\documentclass[conference]{IEEEtran}
\usepackage{amsmath,amssymb,amsfonts}
\usepackage{algorithmic}
\usepackage{graphicx}
\usepackage{textcomp}
\usepackage{xcolor}
\usepackage[backend=biber]{biblatex}
\usepackage{tikz}
\usepackage{circuitikz}
\usepackage{pgfplots}
\pgfplotsset{compat=1.18}
\usepackage{hyperref}
\usepackage{breakurl}
\begin{document}

\title{Efficacy of Full-Packet Encryption in Mitigating Protocol Detection for Evasive Virtual Private Networks}

\author{\IEEEauthorblockN{Amy I. Parker}
\IEEEauthorblockA{\textit{Department of Computer Science} \\
\textit{California State University, Fullerton}\\
Fullerton, CA, USA \\
amyipdev@csu.fullerton.edu}}

\maketitle

\begin{abstract}
Full-packet encryption is a technique used by modern evasive Virtual Private Networks (VPNs) to avoid protocol-based flagging from censorship models by disguising their traffic as random noise on the network. Traditional methods for censoring full-packet-encryption based VPN protocols requires assuming a substantial amount of collateral damage, as other non-VPN network traffic that appears random will be blocked. I tested several machine learning-based classification models against the Aggressive Circumvention of Censorship (ACC) protocol, a fully-encrypted evasive VPN protocol which merges strategies from a wide variety of currently in-use evasive VPN protocols. My testing found that while ACC was able to survive our models when compared to random noise, it was easily detectable with minimal collateral damage using several different machine learning models when within a stream of regular network traffic. While resistant to the current techniques deployed by nation-state censors, the ACC protocol and other evasive protocols are potentially subject to packet-based protocol identification utilizing similar classification models.
\end{abstract}

\begin{IEEEkeywords}
censorship circumvention, VPNs, protocol obfuscation, machine learning classification
\end{IEEEkeywords}

\section{Introduction} \label{Section_Introduction}
Internet censorship, particularly at the nation-state level, has become a pervasive means of restricting access to information and digital services. Virtual Private Networks (VPNs), which create encrypted tunnels between networks \cite{1625300}, have become a vital tool in evading these censorship regimes \cite{6778916}. Users can send encrypted traffic --- not demonstrably meeting topical censorship criteria --- to a VPN provider outside the censorship regime, which will process the network requests and deliver encrypted responses back to the user.

Many censors attempt to combat the usage of VPNs in evading their censorship regimes by simply blocking all traffic utilizing VPN protocols. Whether through exposed header information or other details trivial to fingerprint, such as repeated values at a fixed or local offset, most VPN protocols are easy to detect; aside from typical header detection, Xue et al. studied methods of fingerprinting the OpenVPN protocol \cite{280012}. At the nation-state level, the Islamic Republic of Iran routinely deploys a protocol whitelister, prohibiting the usage on ports 53, 80, and 443 of any protocols other than DNS, HTTP, and HTTPS \cite{257162}. As it is generally uncommon to need to use a VPN across a censorship barrier for purposes not related to evading censorship (such as accessing servers in a foreign office), and alternative methods exist for these use cases, the effects on non-evasive traffic are minimal. 

VPN applications seeking to circumvent these protocol-level restrictions utilize a newer method of evasion: full-packet encryption (FPE). Instead of just encrypting the original datagrams (payloads), FPE-based protocols encrypt the entire packet, including protocol headers; through the properties of cryptographically secure encryption methods, the packets become indistinguishable from random noise \cite{287260}. Traditional protocol detection systems rely on identifiable headers in packets, but in FPE-based protocols, no such identifiable headers exist, making trivial protocol detection (also referred to as "deterministic" detection) impossible. Censors targeting FPE-based VPNs often instead use probabilistic, or "nondeterministic", models which compute heuristics on packets and censor based on probabilities yielded by those heuristics; China's modern approaches to censoring FPE utilize such nondeterministic methods \cite{287260}.

In this paper, I present a detailed series of tests which utilize machine learning classification systems --- a nondeterministic method --- to attempt to discriminate between packets of a fully-encrypted VPN protocol and, depending on the test, either random noise or regular network traffic. The selected fully-encrypted VPN protocol is the Aggressive Circumvention of Censorship (ACC) protocol \cite{acc-paper}, which provides the \texttt{libacc} library \cite{libacc} for encrypting and decrypting packets outside of a running ACC VPN tunnel. After training a given classification model on both types of packets, test packets are passed to the model for evaluation. I report back on the results of these tests, and identify future areas of research for FPE models.

The rest of the paper is organized as follows. \S\ref{Section_Prior_Work} reviews prior work in detecting and analyzing FPE models. \S\ref{Section_Methodology} describes my methodology, the machine learning classifiers, and the collection and encryption of the sample packets. \S\ref{Section_Results} reviews the results of the tests and discusses these results. Finally, \S\ref{Section_Conclusion} concludes.

\section{Prior Work} \label{Section_Prior_Work}
The most primitive form of censoring FPE-based protocols --- protocol whitelisting --- has been studied at the nation-state level primarily in relation to the Islamic Republic of Iran. The first appearance occurred prior to Iran's June 2013 elections, and was studied by Aryan et al. {\cite{179184}} as part of a broader review of the nation's substantial Internet censorship regime. The censorship models would, upon detecting abnormal behavior of a whitelisted protocol or a non-whitelisted protocol, throttle the user's connection, substantially limiting connectivity and causing larger traffic loads such as file downloads to fail. 

Bock et al. \cite{257162} studied Iran's newer protocol whitelist model in 2020. This model is inverted from typical protocol whitelists; rather than block all traffic on non-whitelisted ports and test for detectable protocols, Iran's new regime allows all traffic not on ports 53, 80, and 443 (DNS, HTTP, and HTTPS, respectively), and only permits the three aforementioned protocols on those ports. This model blocks FPE models that run under another protocol's port, a method which some FPE models use to evade traditional port-based protocol whitelist systems.

FPE-based protocols that have traditionally been able to circumvent these methods, such as Shadowsocks, VMess, Obfs4, and Outline, were temporarily blocked by the People's Republic of China's Great Firewall (GFW) in November 2021, as analyzed by Wu et al. \cite{287260}. In addition to checking for protocol fingerprints and sequences of readable characters, the 2021 GFW added a new criterion for blocking packets. If all other whitelisting tests failed, packets would be blocked if the average number of set bits per byte (popcount) was between 3.4 and 4.6. This method attempts to use the apparent randomness of full-packet encryption models against the models, as random packets are likely to have a popcount very close to 4.0. 

As the 2013 and 2020 Iranian whitelist models can be overcome by adapting the design of a protocol, I do not include their models in tests. Likewise, strategies identified by Wu et al., as well as trivial strategies such as randomly varying nonces or pre-encryption padding to trigger a passing popcount, are able to overcome the 2021 GFW; I thus do not include these models in the tests.

\section{Methodology} \label{Section_Methodology}

\subsection{Packet Collection}
Packets were collected using \texttt{libpcap} from a Proxmox node operating a public-facing WordPress instance. The collection algorithm accepted all IPv4 and IPv6 packets except for ARP requests, IPv4 UDP packets (mostly used for node-to-node \texttt{corosync} within the Proxmox cluster and thus not representative of regular network traffic). and ICMPv6 Router Solicitations and Router Advertisements. Requests to local network resources, such as MariaDB instances and the instance's NGINX reverse proxy, were included in the capture; while these are not traditionally accessed by end-users over a VPN, the protocols are not significantly different, and the packets overall make up a minimal percentage of captured traffic. Packets longer than the ACC VPN's Maximum Transmission Unit (MTU) size of 1280 bytes were also dropped.

This collection returned approximately four million packets over three weeks. Due to the limits of available hardware research, this amount was too great to process, making data reduction necessary. 87.5 percent of the packets were randomly eliminated from the collection, reducing the dataset to 474,619 packets totaling approximately 69.6 MiB of data.

\subsection{ACC Encryption}

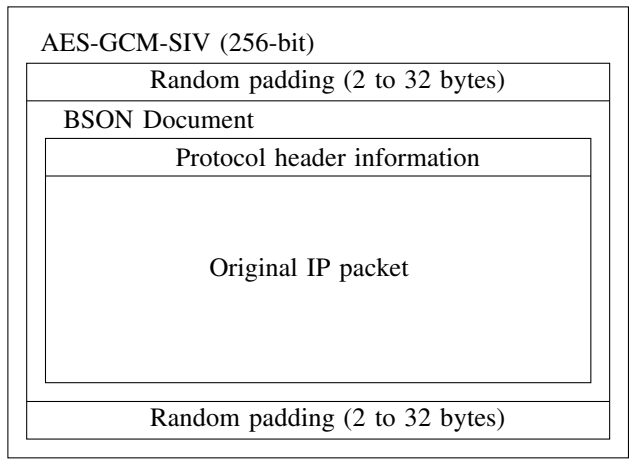
\begin{figure}[!ht]
\centering
\begin{circuitikz}
\tikzstyle{every node}=[font=\normalsize]
\draw  (3.75,33.5) rectangle (12,27.5);
\node [font=\normalsize] at (6,33.75) {VPN packet sent over network};
\draw  (4,32.75) rectangle (11.75,27.75);
\node [font=\normalsize] at (6,33) {AES-GCM-SIV (256-bit)};
\draw [short] (4,32.25) -- (11.75,32.25);
\node [font=\normalsize] at (8,32.5) {Random padding (2 to 32 bytes)};
\draw [short] (4,28.25) -- (11.75,28.25);
\node [font=\normalsize] at (8,28) {Random padding (2 to 32 bytes)};
\node [font=\normalsize] at (5.75,32) {BSON Document};
\draw  (4.25,31.75) rectangle (11.5,28.5);
\draw [short] (4.25,31.25) -- (11.5,31.25);
\node [font=\normalsize] at (8,31.5) {Protocol header information};
\node [font=\normalsize] at (7.75,30) {Original IP packet};
\end{circuitikz}
\label{fig:my_label}
\caption{Structure of an ACC packet}
\end{figure}

Collected packets were then processed through a script utilizing \texttt{libacc} \cite{libacc} to generate equivalent packets to those that would be sent over the network during an ACC connection. ACC uses a three-step encapsulation model to prepare packets for transmission over the network. The original IP packet is encapsulated into a Binary JavaScript Object Notation (BSON) document alongside protocol header information. In the second phase, a variable amount of random padding is placed along with character delimiters ('{' and '}') on either side of the BSON document. Finally, the packet is encrypted with the GCM-SIV mode of AES, which increases resistance against nonce collisions \cite{10.1007/978-3-319-78381-9_18}; the nonce is appended to the end of the packet. These packets are then written out to a separate directory from the original packets.

ACC was selected as the protocol for testing due to its theoretical equivalence in results with other protocols, the combination of many existing protocols' strategies in the development of ACC, and the ease of integrating it into the testing pipeline. ACC is not dependent on pre-existing application states, other than the registration of an AES key with the target server, making encryption of individual packets easier. \texttt{libacc} also provides a much easier interface than other VPN systems (such as Shadowsocks and Obfs4) for encrypting the packets.

\subsection{Random Packet Generation}
Random packets were generated for the tests comparing ACC packets to random data. To avoid packet-size-based influence on these tests, the distribution of packet size in the random packets was set to be the same as the post-processing ACC packets. A total of 474,619 random packets (107.7 MiB) were generated using Python's \texttt{os.urandom()} interface, which pulls from the \texttt{/dev/urandom} character device on Linux systems.

\subsection{Classification Models}
All algorithms were implemented using either \texttt{scikit-learn} 1.5.2 or TensorFlow 2.7.10 on Python 3.11.2 as shipped with Debian GNU/Linux 12. For applicable models, all sixteen threads of the research server's CPU were assigned a worker. Eight machine learning classification algorithms were used: 
\begin{itemize}
\item Quinlan's \textbf{C4.5} \cite{quinlan2014c4} is a simple decision tree classification algorithm that runs in a single step with minimal parameters, implemented using \texttt{sklearn.tree.DecisionTreeClassifier} with no additional parameters.
\item $k$-Nearest Neighbors (\textbf{$k$-NN}) classifies objects based on the classification of the $k$ nearest neighbors to an object. \texttt{sklearn.neighbors.KNeighborsClassifier} was used with \texttt{n\_neighbors} ($k$) set to $\sqrt{len(data_{training})}$.
\item Multilayer Perceptron (\textbf{MLP}) is a standard feedforward neural network model. Two different MLP implementations were used:
    \begin{itemize}
    \item ACC versus random tests used TensorFlow with a Keras model of form [Dense(950), Dense(950; RELU), Dense(950; RELU), Dense(1; Sigmoid)].
    \item ACC versus unencrypted network traffic tests used \texttt{sklearn.neural\_network.MLPClassifier} with hidden layer sizes of $(950, 950, 950)$ and a tolerance of $10^{-9}$.
    \end{itemize}
\item \textbf{Random Forest}, which was implemented through \texttt{sklearn.ensemble.RandomForestClassifier}, combines multiple decision trees for classification; the number of trees, \texttt{n\_neighbors}, was set to 128.
\item \textbf{Logistic Regression} attempts to map an input dataset to an output dataset by configuring parameters of a logistic function; while not as useful for these categories, it sets a minimum baseline for performance. Implementation was done via \texttt{sklearn.linear\_model.LogisticRegression} with a maximum of 10,000 iterations and a tolerance of $10^{-6}$.
\item Convolutional Neural Networks (\textbf{CNNs}) transform a dataset with a series of convolutions which can be configured to develop a model before finally outputting a classification result. While typically used for images and multi-dimensional data, CNNs can be used for linear data, albeit uneffectively when such data is grouped into blocks such as with an AES encryption pass. A CNN model was implemented with Keras layers of [Conv1D(1500, 32; RELU), MaxPooling1D((5,)), Conv1D(128, 32; RELU), MaxPooling1D((5,)), Conv1D(128, 32; RELU), MaxPooling1D((5,), Flatten, Dense(128; RELU), Dense(1; Sigmoid)].
\item Support Vector Machines (\textbf{SVMs}) use linear models along with tunable kernel functions to classify data. The SVM model was implemented with \texttt{sklearn.linear\_model.SGDClassifier}, the \texttt{hinge} loss function (which implements a linear SVM), and a tolerance of $10^{-8}$.
\item Recurrent Neural Networks (\textbf{RNNs}) use recurrent units, which hold a hidden state during training and update based on both the previous hidden state and the current input. The RNN was implemented with a Keras model of form [Embedding(1500, 1500), Bidirectional(LSTM(750)), Dense(1; Sigmoid)].
\end{itemize}

All eight algorithms were used for the ACC versus random data tests. Due to the incredibly significant results exhibited, only C4.5, $k$-NN, and MLP were used for testing ACC packets against regular network traffic. Input data was provided to the models in the form of NumPy arrays of shape $(1500,1,dtype=uint8)$. TensorFlow models used the \texttt{adam} optimizer and \texttt{BinaryCrossentropy} loss function with \texttt{from\_logits=True}.

For the ACC versus real network data tests in particular, IPv4 and IPv6 headers, as well as TCP and UDP headers, were stripped from the packets when comparing them to the ACC packets. In transit, the ACC packets would have IP and transport headers, but \texttt{libacc} does not yield these; as such, a model could easily detect the presence of network and transport layer headers, ruining the tests. An alternative and nearly equivalent method would have been to add IP and TCP/UDP headers to the ACC packets; eliminating them reduces testing variability. 

\subsection{Metrics}

Both \texttt{scikit-learn} and TensorFlow models output five primary metrics used for evaluating and comparing the efficacy of models:
\begin{itemize}
\item \textbf{Accuracy} measures the proportion of test results which are correct, being defined as
\[ A = \frac{TP+FP}{TP+FP+TN+FN} \]
\item \textbf{Precision} measures the proportion of true positives among all positives, being the true positive rate; it is defined as
\[ P = \frac{TP}{TP+FP} \]
\item \textbf{Recall} measures the ability to detect positives in the input by measuring the proportion of ACC packets that were detected, defined as
\[ R = \frac{TP}{TP + FN} \]
\item \textbf{Collateral Damage} measures the percentage of packets which were unintentionally censored. For censors, keeping collateral damage low is important to ensure that networks are still reasonably accessible by compliant users \cite{287260}. Unlike all other metrics measured here, a \textit{lower} collateral damage score is better. It is defined as
\[ C = \frac{FP}{FP + TN} \]
\item \textbf{$F_1$-score} provides a generalized metric for comparing classifiers. It is the harmonic mean of precision and recall, defined as:
\[ F_1 = \frac{2}{R^{-1}+P^{-1}} = \frac{2(TP)}{2(TP)+FP+FN} \]
\end{itemize}

Note that TP refers to the count of true positives, TN true negatives, FP false positives, and FN false negatives. Metrics were manually calculated from the raw output counts for \texttt{scikit-learn}; for TensorFlow, $F_1$-score and collateral damage were calculated using the precision and recall results, as Keras' $F_1$ metrics were inaccurate and there is no native support for collateral damage.

A censorship model is deemed \textit{effective} at blocking ACC traffic in a test if the following conditions are true:
\begin{itemize}
\item $F_1 > 0.95$; this is a common criterion for any classification algorithm, as it accounts for the effects on both ACC and non-ACC traffic better than accuracy and other "complete" measures.
\item $C < 0.01$; if more than 1\% of data is falsely censored, the impact on the network will be significant, and the model is generally infeasible to deploy. China's temporary block on FPE-based models had a collateral damage rate of 0.6\%, which was only feasible for a short amount of time and later discontinued \cite{287260}.
\end{itemize}

\section{Results} \label{Section_Results}

\subsection{ACC versus Random Packets}
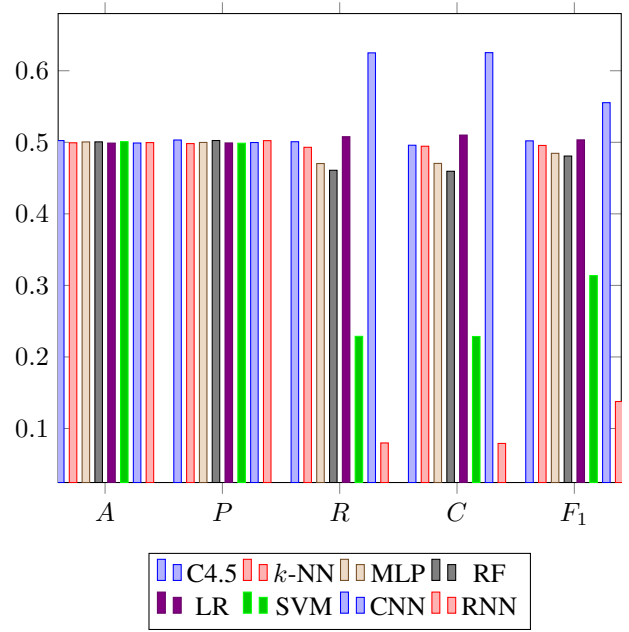
\begin{figure}
    \centering
    \begin{tikzpicture}
        \begin{axis}[ybar, bar width=0.1cm, symbolic x coords={$A$,$P$,$R$,$C$,$F_1$}, xtick=data, width=.5\textwidth, legend style={at={(0.5,-0.15)}, anchor=north, legend columns=4}]
            \addplot table[x=metric,y=c45]{\accvsrandomdata};
            \addplot table[x=metric,y=knn]{\accvsrandomdata};
            \addplot table[x=metric,y=mlp]{\accvsrandomdata};
            \addplot table[x=metric,y=raf]{\accvsrandomdata};
            \addplot table[x=metric,y=logit]{\accvsrandomdata};
            \addplot table[x=metric,y=svm]{\accvsrandomdata};
            \addplot table[x=metric,y=cnn]{\accvsrandomdata};
            \addplot table[x=metric,y=rnn]{\accvsrandomdata};
            \legend{C4.5, $k$-NN, MLP, RF, LR, SVM, CNN, RNN}
        \end{axis}
    \end{tikzpicture}
    \caption{Metrics from ACC versus Random testing}
    \label{fig:rand-graph}
\end{figure}

The results of these tests are displayed in Fig.\, \ref{fig:rand-graph}. All models performed equivalently in the accuracy and precision benchmarks. The CNN-based model overperformed in the $F_1$-score and recall metrics, suffering later in collateral damage. The SVM-based and RNN-based models did the reverse, having low recall and $F_1$-scores but minimal collateral damage; this appears to be a result of the CNN model being too eager (more positive guesses than negative guesses) while the SVM and RNN models were too hesitant (more negative guesses than positive guesses). All other models performed nearly equally ($\sim50\%$) on all other benchmarks.

None of the models were effective in classifying packets as ACC versus non-ACC. Taking into account the eagerness of the CNN-based model and the hesitant nature of the SVM-based and RNN-based models, the classifications returned by each model were essentially random.  

\pgfplotstableread[row sep=\\,col sep=&]{
    metric & c45 & knn & mlp & raf & logit & svm & cnn & rnn \\
    $A$ & 0.5024 & 0.4993 & 0.5005 & 0.5006 & 0.4989 & 0.5009 & 0.4990 & 0.4996 \\
    $P$ & 0.5032 & 0.4982 & 0.4998 & 0.5025 & 0.4991 & 0.4987 & 0.4997 & 0.5023 \\
    $R$ & 0.5008 & 0.4930 & 0.4703 & 0.4609 & 0.5078 & 0.2288 & 0.6250 & 0.0798 \\
    $C$ & 0.4959 & 0.4945 & 0.4705 & 0.4595 & 0.5101 & 0.2286 & 0.6253 & 0.0791 \\
    $F_1$ & 0.5020 & 0.4956 & 0.4846 & 0.4808 & 0.5034 & 0.3137 & 0.5554 & 0.1377 \\
}\accvsrandomdata

\subsection{ACC versus Network Packets}
\begin{figure}
    \centering
    \begin{tikzpicture}
        \begin{axis}[ybar, ytick distance=0.1, ymax=1.05, ymin=0, bar width=0.3cm, symbolic x coords={$A$,$P$,$R$,$10C$,$F_1$}, xtick=data, width=.5\textwidth, legend style={at={(0.5,-0.15)}, anchor=north, legend columns=-1}]
            \addplot table[x=metric,y=c45]{\accvsnetdata};
            \addplot table[x=metric,y=knn]{\accvsnetdata};
            \addplot table[x=metric,y=mlp]{\accvsnetdata};
            \legend{C4.5, $k$-NN, MLP};
        \end{axis}
    \end{tikzpicture}
    \caption{Metrics from ACC vs Network testing}
    \label{fig:net-graph}
\end{figure}
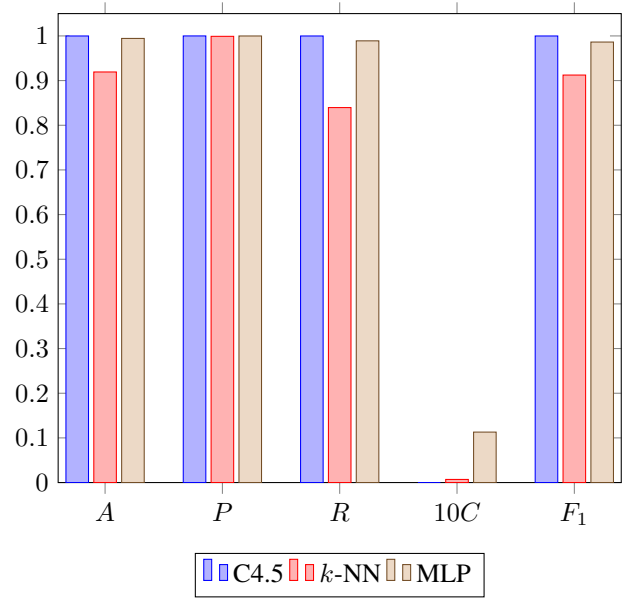

The results of these tests are displayed in Fig.\, \ref{fig:net-graph}; note that $C$ has been scaled by a factor of $10$ for ease of visibility.

\pgfplotstableread[row sep=\\, col sep=&]{
    metric & c45 & knn & mlp \\
    $A$ & 0.9999 & 0.9193 & 0.9944 \\
    $P$ & 1.0 & 0.9991 & 0.9999 \\
    $R$ & 0.9999 & 0.8396 & 0.9890 \\
    $10C$ & 0.0 & 0.007 & 0.113 \\
    $F_1$ & 0.9999 & 0.9124 & 0.9863 \\
}\accvsnetdata

The C4.5 model performed almost perfectly. The iteration showed in the graph had only \textit{one} false negative, with every other prediction being correct; the testing set size was 78,925. Re-running this test on the side several times produced similar results, with no false positives and the count of false negatives being within $[0,7]$. While this model is still nondeterministic, as C4.5 is a probabilistic classification algorithm, this behavior makes it nearly equivalent to a deterministic algorithm. Given that it meets the effectiveness criteria ($F_1 \approx 0.9999873 > 0.95$, $C = 0.0 < 0.01$), C4.5 can be clearly classified as effective in identifying ACC packets within a network stream.

$k$-NN did not perform as well, but still exceedingly better than in the ACC-versus-Random tests. It turned out to be very hesitant, having 200 times the number of false negatives as false positives; this was heavily reflected in its $F_1$-score of approximately $0.91239$, making it not effective as a censorship model. However, this hesitancy was able to keep its collateral damage score low at approximately $0.079\%$.

MLP performed very well, reaching a satisfactory $F_1$-score of approximately $0.9863$. While not nearly as effective as C4.5, it achieved impressive results, even with a relatively small model. The particular model trained is not quite effective, as the collateral damage of $1.13\%$ is above the efficacy threshold of $C < 0.01$; however, alterations to the model or an increase in training data quantities could yield better results.

\subsection{Applicability to Censorship Models}
Current individual-packet censorship models utilize deterministic or lightweight nondeterministic tests due to the computational overhead and latency added by censorship models to packet transmission. While checking header information or calculating a packet's popcount requires a minimal amount of additional computation relative to route calculation, deploying a machine learning model, even if pre-trained, adds a significant amount of overhead which current censorship and routing equipment may not be prepared for, requiring expensive investment in network infrastructure that may not be feasible at the nation-state level.

Despite this, the high performance of C4.5 indicates that these models may be deployable now or in the near future. Of the eight algorithms tested, C4.5 is the least computationally intensive; a single thread on a Ryzen 7 3700X (an AMD Zen 2 CPU) was able to achieve $548044.96$ packets classified per second using C4.5 in preliminary testing, which would allow servicing of roughly 8.8 million packets/second across all threads; specialized hardware would likely be able to achieve much higher performance with at reasonably low electrical expenditure and initial cost.

\section{Conclusions and Future Work} \label{Section_Conclusion}

In this work, I developed methods for testing the efficacy of the ACC protocol, which uses the full-packet encryption model, against machine learning classification-based censorship models. While ACC was able to withstand tests against random data, ACC packets were easily identified when in a stream of regular network packets. C4.5 was identified as the most effective model for detecting ACC experimentally, and the ability for C4.5 models to be integrated into censorship middleware was explored.

These tests were only run on the ACC protocol; while all FPE-based protocols should have similar results to ACC, future research should verify that these results apply to other protocols such as Shadowsocks and Obfs4. Additional research can be conducted on identifying which FPE-based protocol is in use, which may identify other detection strategies for these protocols. 

Lastly, the efficacy of masquerade modes (such as sending VPN traffic over HTTPS) should be further investigated, as such methods have previously been the most effective in evading protocol-targeting censorship. TLS-based masquerading remains the most promising; for lower-throughput connections on networks which do not enforce single-DNS servers or block ICMP, tunneling over DNS or ICMP data may introduce new effective strategies.

\section*{Acknowledgment}
I thank my research advisor, Marc Velasco, for the incredible guidance and advice he has provided in bringing this research to conclusion. I also thank the other members of the ACC team --- Kevin Liu, Owen de Vita, Amal Suresh, Reyhanullah Shirzad, and Mason Chiang --- for developing the ACC protocol used in this research.

\printbibliography

\end{document}